# Two – Band Model Interpretation of the *p*- to *n*- Transition in Ternary Tetradymite Topological Insulators

T. C. Chasapis[1*], D. Koumoulis[2], B. Leung[2], N. P. Calta[1], S.-H Lo[3], V. P. Dravid[3], L.-S. Bouchard[2,4] and M. G. Kanatzidis[1*]

[1] Department of Chemistry, Northwestern University, 2145 Sheridan Rd., Evanston, IL, 60208, USA.

[2] Department of Chemistry and Biochemistry, University of California – Los Angeles, 607 Charles E. Young Drive East, Los Angeles, CA, 90095, USA.

[3] Department of Materials Science and Engineering, Northwestern University, 2145 Sheridan Rd., Evanston, IL, 60208, USA.

[4] California NanoSystems Institute – UCLA, 570 Westwood Plaza, Los Angeles, CA 90095, USA.

*Corresponding authors. E-mail address: t-chasapis@northwestern.edu
m-kanatzidis@northwestern.edu

**Abstract:** The requirement for large bulk resistivity in topological insulators has led to the design of complex ternary and quaternary phases with balanced donor and acceptor levels. A common feature of the optimized phases is that they lie close to the *p*- to *n*- transition. The tetradymite $Bi_2Te_{3-x}Se_x$ system exhibits minimum bulk conductance at the ordered composition $Bi_2Te_2Se$. By combining local and integral measurements of the density of states, we find that the point of minimum electrical conductivity at $x=1.0$ where carriers change from hole-like to electron-like is characterized by conductivity of the mixed type. Our experimental findings, which are interpreted within the framework of a two–band model for the different carrier types, indicate that the mixed state originates from different type of native defects that strongly compensate at the crossover point.

## 1. Introduction

Bismuth selenide (Bi$_2$Se$_3$) and bismuth telluride (Bi$_2$Te$_3$) crystallize in the primitive rhombohedral structure formed by a sequel of atomic layers in the form Q(1)–Bi–Q(2)–Bi–Q(1) (Q = Te or Se). In recent years, they have been classified as topological insulators (TI) [1-4]. Unfortunately, the presence of significant bulk conduction poses a challenge for the observation of surface states, leading to significant efforts aimed at suppressing the bulk conductance. Appropriate doping or alloying can lead to bulk insulating behavior. Large bulk resistivities have been reported for Bi$_2$Te$_2$Se [5,6], Bi$_{1.5}$Sb$_{0.5}$Te$_{1.3}$Se$_{1.7}$ [7] and Bi$_{1.4}$Sb$_{0.6}$Te$_{1.8}$S$_{1.2}$ [8]. All of these compositions are located close to a $p$- to $n$- crossover of conductivity. The crossover in the conductivity type is the result of a counteracting effect between different types of defects that arise upon changing the stoichiometry. A crossover point can be attained in multiple ways. For the Bi$_2$Te$_{3-x}$Se$_x$ system a $p$- to $n$- transition is expected close to $x=1.0$ [9]. For Sb$_x$Bi$_{2-x}$Te$_{3-y}$Se$_y$ the $p$- to $n$- transition may be obtained by tuning either $x$ or $y$ [10]. For the Bi$_2$Te$_{3-x}$S$_x$ system the crossover is located at $x\sim0.12$ [11]. Alloying Bi$_2$Te$_2$S with Sb shifts the system closer to the crossover [8].

In this report, we examine the $p$- to $n$- transition of the Bi$_2$Te$_{3-x}$Se$_x$ system more closely by combining results from electrical transport and nuclear magnetic resonance (NMR) measurements. The Seebeck coefficient is positive and increases up to $x=0.6$, due to a reduction in the hole density, which is reflected in a drop in electrical conductivity. For $x \geq 1.4$ the Seebeck coefficient is negative and decreases, reflecting an increase in the electron density, leading to higher electrical conductivity. The compositions with $x<0.8$ and $x \geq 1.4$ are nearly extrinsic and the transport properties are described on the assumption of a single carrier type. The chemical potential of holes was found close to the valence band edge. On the other hand, the chemical potential of electrons was found within the conduction band revealing the metallic behavior of the Se-rich compositions. For $0.8<x<1.0$ the Seebeck coefficient changes sign, the electrical conductivity displays a minimum and the optical energy gap is maximized. The behavior is qualitatively explained within the framework of a two–band model allowing for the overlap of two types of carriers with opposite sign. Based on this two–band model the total Seebeck coefficient changes sign for chemical potentials located within the gap yielding intrinsic conduction. At the same time the electrical conductivity displays a minimum due to minimum carrier density. The combination of the Seebeck and NMR analysis unveils a mixed conduction state across the entire compositional range as well as a decrease of the Fermi level density of states at the ordered composition $Bi_2Te_2Se$. The model captures the general features of the *p*- to *n*- transition, explaining the large bulk resistivity obtained for properly alloyed ternary and quaternary TIs.

## 2. Materials and methods

$Bi_2Te_{3-x}Se_x$ ingots were synthesized by mixing the appropriate ratios of high purity (99.999%) elemental Bi, Te, and Se metals. Starting materials were sealed under high vacuum (~$10^{-4}$ Torr) in quartz tubes, heated to 800 °C for 15 hours, kept at 800 °C for 10 hours, and cooled to room temperature for at least 15 hours. Powder X-ray diffraction (PXRD) measurements were performed using a CPS120 INEL powder X-ray diffractometer (Cu Kα, 1.54056A°). Transmission Electron Microscopy (TEM) images were collected using a JEOL 2100F transmission electron microscope operating at 200 kV. TEM samples were prepared by cryogenic crushing to minimize sample preparation artifacts. Scanning Electron Microscopy (SEM) imaging and Energy Dispersive Spectroscopy (EDS) were performed with a Hitachi S-3400 scanning electron microscope equipped with a PGT energy-dispersive X-ray analyzer. Data were acquired with an accelerating voltage of 20 kV. The room temperature electrical conductivity and Seebeck coefficient were measured for rectangular shaped samples in helium atmosphere on a ULVAC-RIKO ZEM-3 instrument. The measurement direction coincides with the in-plane direction of the hexagonal structure. Room temperature diffuse reflectance measurements of finely ground powders were collected using a Nicolet 6700 FT-IR spectrometer. The direct energy gaps were obtained by extrapolating to zero energy the squared Kubelka – Munk function $(\alpha/s)^2$ relating the diffuse reflectance with the absorption coefficient $\alpha$ and the scattering coefficient *s* [12].

$^{125}$Te NMR spin-lattice relaxation times ($T_1$) and spectra were acquired using a Bruker DSX-300 spectrometer. A standard Bruker X-nucleus probe with 5 mm solenoid coil was used. To avoid skin depth issues of the *RF* transmission power, ingots were ground by mortar and pestle. Spectral data were acquired using a spin-echo sequence. $T_1$ data were acquired with a saturation-recovery technique. The $^{125}$Te chemical shift scale was calibrated using the unified $\varXi$ scale [13, 14].

## 3. Results and discussion

Figure 1a displays the PXRD patterns of selected compositions. All reflection peaks are assigned to the hexagonal crystal structure. A high magnification TEM image at a grain boundary of the $Bi_2Te_2Se$ composition is shown in Fig. 1b. The corresponding selected area electron diffraction pattern confirms that we have a single-phase material with hexagonal crystal structure. The layered crystal structure is further supported by the SEM image of cleaved $Bi_2Te_2Se$ (Fig. 1c). EDS results obtained from different locations of the rectangular shaped $x=1$ sample are summarized in Table I and suggest that the overall composition lies close to the nominal one. The small fluctuations around the $Bi_2Te_2Se$ composition do not suggest phase segregation as evidenced by both TEM and SEM observations but rather a structure that is not fully ordered.

The evolution of the direct energy gap is presented in Fig. 1d: for $x$ values up to 1.0 the band gap increases reaching a maximum of ~0.26 eV, while for higher Se content the band gap remains nearly constant. The increase of the energy gap in $Bi_2Te_{3-x}Se_x$ alloys up to $x=1$ is the result of the substitution of the $Te^{(2)}$ sites by the more electronegative selenium atoms. In $Bi_2Te_3$ the $Bi$–$Te^{(2)}$ bond is weaker than the $Bi$–$Te^{(1)}$ bond and hence determines the energy gap. The substitution of the $Te^{(2)}$ sites by selenium atoms increases the bond strength and hence the energy gap until $x\sim 1$ [15]. The observed increase of the direct energy gap up to $x=1$ may also be explained by a possible crossover of different valleys within the conduction or valence bands [16]. Optical studies of the $Bi_2Te_{3-x}Se_x$ system have shown that for $x=1.0$ a value of ~0.3 eV is expected for the direct energy gap [15, 17]. The slightly lower value in our case may be due either to incomplete ordering or the formation of impurity bands caused by native defects [5, 6].

The saturation of the energy gap for $x>1.0$ (Fig. 1d) is the result of the Burstein effect [15, 18]. For $x>1.4$ the Fermi level is located within the conduction band. Under these conditions $E_{g,opt} = E_g + E_F$ where $E_g$ is the direct gap of the undoped material [18]. In general, a reduction of the energy gap is expected for $x>1.0$, assuming that as selenium atoms go to the $Te^{(1)}$ sites they tend to attract charge along the Bi–$Te_xSe_{1-x}$ axis, making the Bi-$Se^{(2)}$ bond less ionic [15, 19], although the effect of the complex band structure cannot be excluded [16]. This reduction is not seen in our case due to the Burstein effect.

The variation of the room temperature Seebeck coefficient versus Se content is presented in Fig. 2a: the Seebeck coefficient initially is positive and increases, reaching a maximum of ~240 µV/K for $x$=0.6. For Se contents 0.6<$x$<1.4 the Seebeck coefficient decreases. At $x$~0.9 the Seebeck coefficient passes through zero and changes sign for higher $x$ values, denoting the change in conductivity type. For $x$=1.4 the Seebeck coefficient reaches a maximum ~ −200 µV/K and then decreases to ~ −60 µV/K at $x$~2.8. The compositional variation of the room temperature electrical conductivity is presented in Fig. 2b. One can see that the electrical conductivity initially decreases, reaching a minimum at $x$~1.0 and increases for higher Se amounts.

Assuming a single type of carriers and based on the semi-classical Mott−Jones formula the Seebeck coefficient is inversely proportional to the carrier density [20]. This means that for $x$≤0.6 the increase of the Seebeck coefficient (Fig. 2a) captures the decrease in the hole density, i.e. the reduction of the electrical conductivity (Fig. 2b). On the other hand, the reduction of the Seebeck coefficient for $x$>1.4 demonstrates the increase of the electron density followed by a subsequent increase in the electrical conductivity.

Accounting for acoustic phonon scattering and band parabolicity, the Seebeck coefficient as a function of the reduced chemical potential $\eta = E_F / k_B T$ ($E_F$, Fermi level) is given by [21]

$$S = \pm \frac{k_B}{e}\left(\frac{F_{r+2}(\eta)}{F_{r+1}(\eta)} - \eta\right) \quad (1)$$

where $F$ is the Fermi integral which, in the general case with index $r$, is given

$$F_r(\eta) = \int_0^\infty \left(-\frac{\partial f}{\partial \varepsilon}\right)\varepsilon^r d\varepsilon \quad (2)$$

where $f$ is the carrier distribution function. For acoustic phonon scattering $r = 0$ [21]. We note that the assumption of pure acoustic phonon scattering may not fully hold, especially for $Bi_2Se_3$-rich compositions. Transport studies on $Bi_2Se_3$ have shown mixed scattering by both acoustic phonons and ionized impurities. The ionized impurities are due to Se vacancies [22].

The application of Eq. (1) to the $x$=0.4, 0.6, 1.4, 2, 2.5 and 2.8 compositions yielded the reduced chemical potential of holes and electrons as a function of $x$ (Fig. 2a inset). For Fermi levels within the bands $\eta$>0 and $\eta$<0 for Fermi levels in the energy gap. The reduced chemical potential is almost zero for $x$=0.4 and negative for $x$=0.6, while it is positive for Se-rich compositions. These positive values cause the saturation of the experimental energy gap (Fig. 2d). We therefore conclude that upon alloying the Fermi level moves from the valence band edge for the $Bi_2Te_3$-rich compositions into the conduction band for the $Bi_2Se_3$-rich materials.

For the compositions $x=0.8$ and $x=1.0$ the interpretation of the experimental Seebeck coefficient assuming only one carrier type fails. These materials show Seebeck values close to the $p$- to $n$- crossover. The Seebeck coefficient drops as a function of $\eta$ (Eq. (1)). Low absolute Seebeck values yield Fermi levels within the bands. On the assumption of a single band model, the reduced chemical potential of holes for the $x=0.8$ and electrons for the $x=1$ would be located within the valence and the conduction band respectively. This situation entails large electrical conductivities in disagreement with the experimental results (Fig. 2b). Alternatively, another physical explanation may be obtained by employing a two–band model with a conduction band and a valence band separated by an energy gap and taking into account two types of carriers with opposite sign. The contribution of the positive ($h$) and the negative ($e$) charge carriers to the overall Seebeck coefficient can be described by [23]

$$S = \frac{S_h - \frac{\sigma_e}{\sigma_h}|S_e|}{1+\frac{\sigma_e}{\sigma_h}} \tag{3}$$

where $S_h$ and $S_e$ are the partial Seebeck coefficients of holes and electrons and $\sigma_h$ and $\sigma_e$ are the respective electrical conductivities. The conductivity ratio is a function of the ratio of the effective masses $m^*_{e,h}$ and the ratio of the partial mobilities $\mu_{e,h}$ [23]

$$\frac{\sigma_e}{\sigma_h} = \left(\frac{m^*_e}{m^*_h}\right)^{3/2} \cdot \frac{\mu_e}{\mu_h} \cdot \frac{F_{r+1}(\eta_e)}{F_{r+1}(\eta_h)} = A \cdot \frac{F_{r+1}(\eta_e)}{F_{r+1}(\eta_h)} \tag{4}$$

where $A = \left(\frac{m^*_e}{m^*_h}\right)^{3/2} \cdot \frac{\mu_e}{\mu_h}$ (5)

The reduced chemical potential of holes is related to the reduced chemical potential of electrons through $\eta_h = -\eta_e - E_g/k_BT$ where $E_g$ is the energy gap between the two bands. The total Seebeck values and the belonging reduced Fermi level depend sensitively on the gap width and on the $A$ parameter [23].

The direct application of a two–band model in our case is not straightforward due to the relatively large number of parameters. For $n$-type doped $Bi_2Te_{3-x}Se_x$, studies on the transport properties have shown that the electron effective mass decreases with $x$ [24, 25], while the electron mobility was proposed to remain nearly unchanged up to $x\sim 1$ [25]. The mobility of holes is expected to decrease upon alloying with Se. For both $Bi_2Te_3$ and $Bi_2Se_3$ the valence band structure is complex [26, 27]. Different valleys in both the conduction and the valence bands may contribute to conduction upon alloying [16]. All the above make the estimation of the *A* parameter challenging. Owing to this, this analysis of mixed conduction can be at most qualitative.

Figure 2c displays the variation of the total Seebeck coefficient against reduced Fermi level on the simplified assumption of $A=1.0$ and $E_g\sim0.26$ eV. The theoretical Seebeck coefficient displays two regions of extrinsic conduction ($S=S_h$ and $S=S_e$), onset of mixed conduction $S\sim S_{max}$, and primarily intrinsic conduction $S<<S_{max}$. The variation of the calculated electrical conductivity is shown in Fig. 2d, where a minimum near the *p*- to *n*- crossover is observed as a result of minimum net carrier density.

Focusing on the *x*=0.8 and 1.0 compositions we may conclude that intrinsic conduction define the Fermi level located deep within the energy gap, under the assumptions of $A=1.0$ and energy gap lying close to the optical gap. Intrinsic behavior originates from the interaction of different types of native defects. The *p*-type conductivity in $Bi_2Te_2Se$ originates mostly from $Bi_{Te}$ antisites, while the *n*-type character is due to Te vacancies or $Te_{Bi}$ antisites [28]. Since in our case the *x*=1.0 material is not fully ordered Se vacancies are also a source of *n*-type conduction The EDS results (Table 1) support regions with slightly different Bi/Te ratio that may account for different type of carriers. Under certain conditions, the counteracting effects between native defects may lead to Seebeck values close to the *p*- to *n*- transition and to a minimum of the electrical conductivity.

Based on Fig. 2c it is clear that for a given $E_g$ and *A* the type of conductivity and the intrinsic or extrinsic conditions may be tuned by the Fermi level. The relative contribution of electrons and holes results in Seebeck values ranging from positive to negative. The theoretical dependence of the Seebeck coefficient of Fig. 2c may qualitatively account for the different Seebeck values reported on the $Bi_2Te_2Se$ compound as a result of small variations in stoichiometry. Greenaway *et* al. [15] and Akrap *et* al. [29] reported values of ~ −240 and ~ −300 μV/K respectively indicative of nearly extrinsic *n*-type conduction. On the other hand, Fuccillo *et al.* reported nearly extrinsic *p*-type conduction with Seebeck values as high as ~ 400 μV/K [30]. On the contrary, LaChance *et al*. reported Seebeck values close to zero and intrinsic conduction [31]. The similarity of the experimental and theoretical behaviors of Fig. 2 suggests that a two–band model may describe the whole compositional dependence of the transport properties. For a given Fermi level, band width and conductivity ratio one expects the Seebeck coefficient for each *x* value to be described by the theoretical curve of Fig. 2c.

The aforementioned interaction of the *n*-type defects with the *p*–type ones leads to the inaccurate determination of carrier concentration derived from volume-averaged (i.e. integral) techniques such as Seebeck coefficient measurements [15, 29-31]. In contrast to electrical transport methods, the NMR lineshape and $T_1$ measurements allow for a site-specific characterization of the electronic band structure [14] due to their proportionality to the effective mass and the carrier density. We have measured $^{125}$Te NMR spectra and $T_1$ measurements as function of Se content to extract microscopic information related to changes in the density of states at the Fermi-level. We extracted frequency shift and linewidth of each spectrum as a function of Se content. There is a linear dispersion between the $^{125}$Te shift and conductivity, indicating that the Bi$_2$Te$_{3-x}$Se$_x$ system follows a mixed conduction mechanism provided by *p*-type and *n*-type carriers (Fig.3a). Interestingly, this composition dependence of the NMR shift is more pronounced in *x*=1, which exhibits a minimum with a shift difference of 300 ppm from the case of *x*=0.2 and 100 ppm from the *x*=1.8. In addition, the $^{125}$Te NMR linewidth narrows with increasing Se content. Close to *x*=1, the linewidth achieves its lowest value, which we attribute to a minimum in electrical conductivity (Fig. 2b).

The $^{125}$Te magnetization recovery data in Bi$_2$Te$_{3-x}$Se$_x$ series were fitted to a stretched exponential function, as a stretched exponential model describes well materials dominated by structural defects and electronic inhomogeneities [14, 32, 33]. The presence of structural defects within the composition range is consistent with the observed value of the Kohlrausch exponent (*β*) [32]. For Bi$_2$Te$_{3-x}$Se$_x$ *β* was found to be 0.7 [14]. In case of semiconductors such as Bi$_2$Te$_{3-x}$Se$_x$, the spin-lattice relaxation rate is proportional to $(m^*)^{\frac{3}{2}} \cdot n \cdot (k_B \cdot T)^{\frac{1}{2}}$, where $m^*$ is the effective mass of the carriers, $k_B$ is the Boltzmann constant and $n$ is the charge carrier concentration [34-37].

The charge carrier concentration due to the mixed conductive state of Bi$_2$Te$_{3-x}$Se$_x$ originates from both electrons and holes, $n_e$ and $n_h$ that change differently by a thermally activated *T*-dependence [37]. NMR in PbSe [38] and PbTe [39] has probed similar behavior. At ambient temperature, the interplay between $n_e$ and $n_h$ is constant as expressed by the conductivity but strongly varies across the Bi$_2$Te$_{3-x}$Se$_x$ series. In Fig. 3b, we monitored this interplay between electrons and holes by plotting the $T_1^{-1}$ versus the electrical conductivity. The linear dispersion between the relaxation rate and conductivity indicates that the entire system follows a mixed conduction mechanism analogous to the aforementioned $^{125}$Te shift behavior. Hence counteracting effects between the hole-rich, and electron-rich, conductivity varies systematically. To obtain a clear view of the electronic changes in the vicinity of the *p*-to-*n* transition, we plotted the spin-lattice relaxation time, the Seebeck coefficient and the carrier-density product $1/(T_1 \cdot T^{1/2})$ versus Se content (Fig. 3c). The gray region indicates an *n*-type phase, whereas the white region shows a *p*-type phase. The larger $T_1$ value at the *p*- to *n*- crossover is consistent with a decreased net carrier concentration in the *x*=1.0 composition. In addition, the quantity $1/(T_1 \cdot T^{1/2})$ which is proportional to the charge carrier concentration (*N*) [14,35-37], achieves a minimum value at the *x*=1.0 composition. All the above results suggest that counteracting processes between defects modify the electronic characteristics. This counter-effect could be the driving force of the gradual transition from the *p*-type to *n*-type conductivity in $Bi_2Te_{3-x}Se_x$.

## 4. Conclusion

The *p*- to *n*- transition of the $Bi_2Te_{3-x}Se_x$ series was monitored via room temperature Seebeck coefficient and NMR spin-lattice relaxation. The progressive substitution of Te by the more electronegative Se atoms changes the local structure, leading to the increase of the direct gap. For *x*<0.8 the positive Seebeck values reflect hole-like conduction mostly due to $Bi_{Te}$ antisite defects. For *x*>1.4 the Seebeck coefficient displays negative values due to *n*-type behavior as a result of Te/Se vacancies. The balance between different types of defects drives the compositional variation of the Seebeck coefficient and $^{125}$Te NMR $T_1^{-1}$ values. For 0.8≤*x*≤1.0 strong counteracting effects between donors and acceptors lead to a situation of mixed carrier conduction. In that region, the Seebeck coefficient changes sign with small absolute values. This transition is interpreted within the framework of a two–band model. The $^{125}$Te NMR resonance frequency, the linewidth and the spin- lattice relaxation time peak at *x*=1. The linear dependence of the $^{125}$Te NMR shift and spin-lattice relaxation rate versus conductivity indicates that the system follows a mixed conduction mechanism. Due to this interaction, minimum electrical conductivity can be defined as the point where the $1/(T_1 \cdot T^{1/2})$ product reaches its lowest value. This product is proportional to the charge carrier density in the vicinity of the *p*- to *n*- crossover.

Therefore, the two-band model captures the general features of the *p-* to *n-* transition and explains the large bulk resistivity obtained for properly alloyed ternaries and quaternaries TIs.

## 5. Acknowledgments

This research was supported by the Defense Advanced Research Project Agency (DARPA), Award No. N66001-12-1-4034. Part of this work made use of the EPIC facility (NUANCE Center-Northwestern University), which has received support from the MRSEC program (NSF DMR-1121262) at the Materials Research Center; the International Institute for Nanotechnology (IIN); and the State of Illinois, through the IIN.

## 6. References

[1] D. Hsieh, Y. Xia, D. Qian, L. Wray, F. Meier, J. H. Di, J. Osterwalder, L. Patthey, A. V. Fedorov, H. Lin, A. Bansil, D. Grauer, Y. S. Hor, R. J. Cava and M. Z. Hasan, *Phys. Rev. Lett.* **103,** 146401 (2009).
[2] M. Z. Hasan and C. L. Kane, *Rev. Mod. Phys.* **82,** 3045 (2010).
[3] Y. Xia, D. Qian, D. Hsieh, L. Wray, A. Pal, H. Lin, A. Bansil, D. Grauer, Y. S. Hor, R. J. Cava and M. Z. Hasan, *Nature Phys.* **5**, 398 (2009).
[4] J. E. Moore and L. Balents, *Phys. Rev. B* **75** 121306(R) (2007).
[5] Z. Ren, A. A. Taskin, S. Sasaki, K. Segawa and Y. Ando, *Phys. Rev. B* **82**, 241306(R) (2010).
[6] S. Jia, H. Ji, E. Climent-Pascual, M. K. Fuccillo, M. E. Charles, J. Xiong, N. P. Ong and R. J. Cava, *Phys. Rev. B* **84**, 235206 (2011).
[7] Z.Ren, A. A.Taskin, S. Sasaki, K. Segawa, and Y. Ando, *Phys. Rev. B* **84**, 165311 (2011).
[8] Ji Huiwen, J. M. Allred, M. K. Fuccillo, M. E. Charles, M. Neupane, L. A. Wray, M. Z. Hasan, and R. J. Cava, *Phys. Rev. B* **85**, 201103(R) (2012).
[9] N. Fuschillo, J. N. Bierly and F. J. Donahoe, *J. Phys. Chem. Solids* **8**, 430 (1959).
[10] I. Teramoto and S. Takayanagi, *J. Phys. Chem. Solids* **19**, 124 (1961).
[11] J.Horak., P.Lostak, L. Koudelka and R. Novotni ,*Solid State Commun.* **55**, 1031 (1985).
[12] B. Philips-Invernizzi, D. Dupont, and C. Caze, *Opt. Eng.* **40**, 1082 (2001).
[13] R. K. Harris, E. D. Becker, S. M. C. De Menezes, R. Goodfellow, P. Granger, *Pure Appl. Chem.* **73**, 1795−1818 (2001).
[14] D. Koumoulis, , T.C. Chasapis, , B. Leung, R.E. Taylor, D.Jr. King, M.G. Kanatzidis, L.S. Bouchard, *Adv. Funct. Mater.* **24**, 1519−1528 (2014).
[15] D.L. Greenaway and Harbeke, *J. Phys. Chem. Solids* **26**, 1585 (1965).
[16] H.K. Köhler W. Haigis, and A von Minddendorf, *Phys. Stat. Solidi b* 78, 637 (1976).
[17] I.G. Austin and A.J. Sheard, *J. Electron. Contr.* **3**, 236 (1957).
[18] Z.M. Gibbs, A. LaLonde and G.J. Snyder, *New Journal of Physics* **15**, 075020 (2013).
[19] J. R. Drabble and C. H. L. Goodman, *J. Phys. Chem. Solids* **5**, 142 (1958).
[20] G. J. Snyder and E. S. Toberer , *Nat. Mater.* **7**, 105 (2008).
[21] B.M. Askerov, *Electron Transport Phenomena in Semiconductors* (World Scientific, Singapore, 1994)

[22] J. Navratil, J. Horák, T. Plechácek, S. Kamba, P. Lośt'ák, J.S. Dyck, W. Chen, and C. Uher, *Journal Solid State Chem.* 177, 1704 (2004).
[23] Z.M. Gibbs, H.-S. Kim, H.Wang, and G.J. Snyder, *Appl. Phys. Lett.* 106, 022112 (2015).
[24] G.N. Gordiakova, G.V. Kokosh and S.S. Sinani, *Sov. Physics-Technical Physics* **3**, 1 (1958).
[25] S.Wang, G. Tan, W. Xie, G. Zheng, H. Li, J. Yang, and X. Tang, *J. Mater. Chem.* **22**, 20943 (2012).
[26] A. Kulbachinskii, M. Inoue, M. Sasaki, H. Negishi, W.X. Gao, K. Takase, Y. Giman, P. Lostak and J. Horak, *Phys. Rev. B* **50**, 16921 (1995).
[27] Yi-Bin Gao, B. He, D. Parker, I. Androulakis, and J.P. Heremans, *Phys. Rev. B,* **90**, 125204 (2014).
[28] D.O. Scanlon, P.D. King, R.P. Singh, A. de la Torre, S.M. Walker, G. Balakrishnan, F. Baumberger and C.R. Catlow, *Adv. Mater.* **24**, 1254 (2012).
[29] A. Akrap, A. Ubaldini, E. Giannini and L. Forró, *EPL* **107** 57008 (2014).
[30] M.K. Fuccilo, J.A. Shuang, M.E. Charles, and R.J. Cava, *J. Electron. Mater.* **42**, 1246 (2013).
[31] M.H. LaChance, and E.E. Gardner, *Adv. Energy Conv.* **1**, 133 (1961).
[32] R. Kohlrausch, *Ann. Phys. Chem.* **91**, 179−213 (1854).
[33] D. C. Johnston, *Phys. Rev. B* **74**,184430 (2006).
[34] C.S.Lue, Y.F. Tao, K.M. Sivakumar, Y.K. Kuo, *J. Phys.:Condens. Matter*, **19**, 406230-8 (2007).
[35] N. Bloembergen, *Physica, Amsterdam* **20**, 1130 (1954).
[36] D. Wolf, *Spin Temperature and Nuclear-Spin Relaxation in Matter*, Clarendon, (Oxford, 1979).
[37] H. Selbach, O. Kanert, D. Wolf, *Phys. Rev. B* **19**, 4435 (1979).
[38] D. Koumoulis, R. E. Taylor, D. Jr. King, L-S. Bouchard, *Phys. Rev. B* **90**, 125201 (2014).
[39] R.E. Taylor, F. Alkan, D. Koumoulis, M.P. Lake, D. Jr. King, C. Dybowski, L.-S. Bouchard, *J. Phys. Chem. C* **117**, 8959 (2013).

TABLE I: EDS results at different locations (spots, regions) of the $x=1$ material.

| Nominal | | $Bi_2Te_2Se$ | |
|---|---|---|---|
| EDS | *Bi* | *Te* | *Se* |
| | 2.0 | 1.8 | 1.2 |
| | 2.0 | 2.3 | 0.7 |
| | 2.0 | 2.2 | 0.8 |
| | 2.0 | 2.0 | 1.0 |
| | 2.0 | 1.8 | 1.2 |
| | 2.0 | 1.9 | 1.1 |
| | 2.0 | 2.1 | 0.9 |
| Average[*] | 2.0±0.10 | 2.0±0.1 | 1.0±0.05 |

[*]*Assuming an instrumental error of ±5%*

**FIGURES**

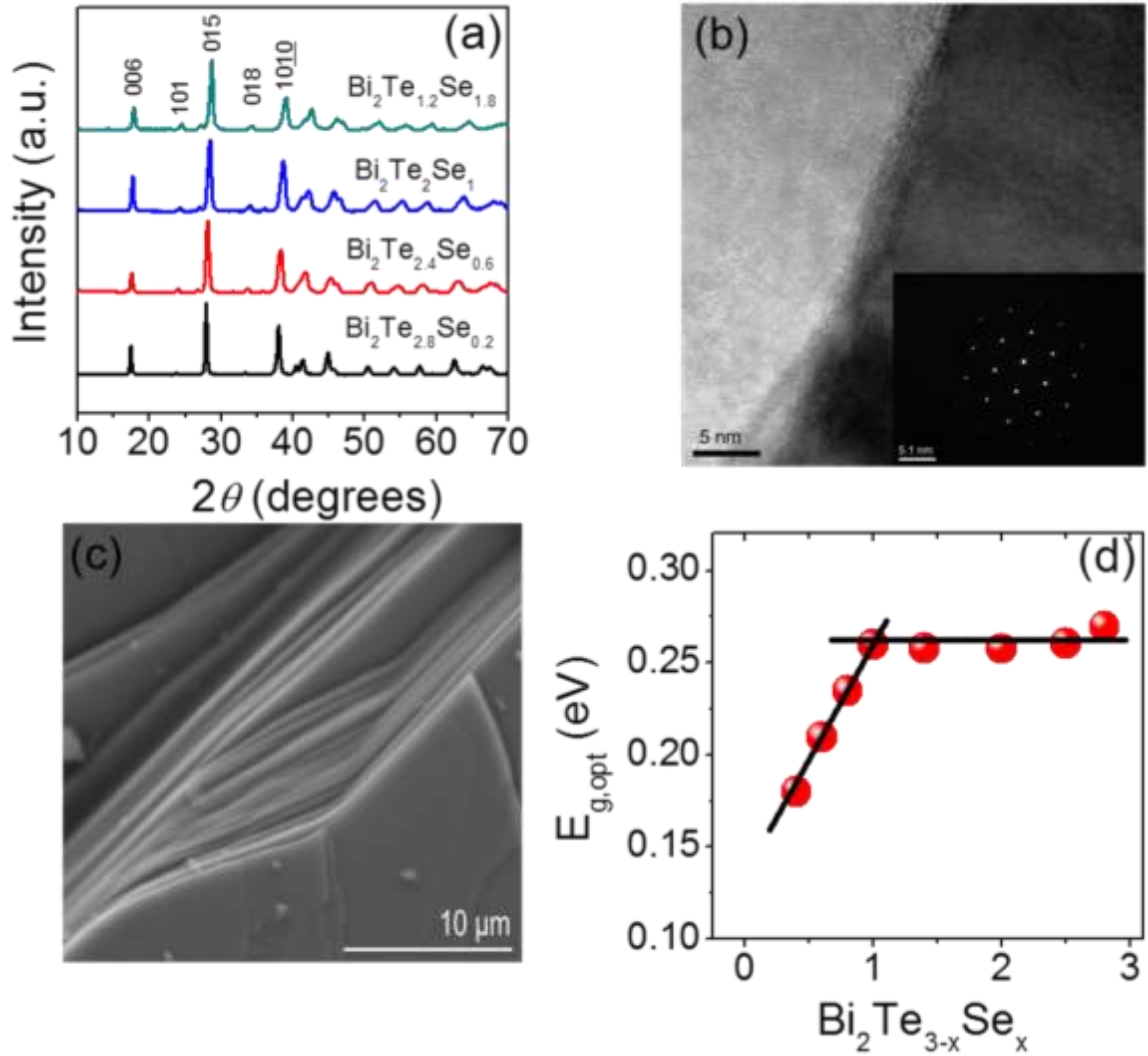

**FIG. 1** (Color online) (a) PXRD patterns of selected compositions, (b) TEM image at a grain boundary of $Bi_2Te_2Se$ composition and the respective representative electron diffraction pattern from one grain oriented along *c*-axis of the unit cell. (c) SEM image of cleaved surface crystal of the $Bi_2Te_2Se$ material, (d) Compositional dependence of the optical direct energy gap. The saturation for x>1 is attributed to the Burstein shift (see text). Solid lines are guide to the eye.

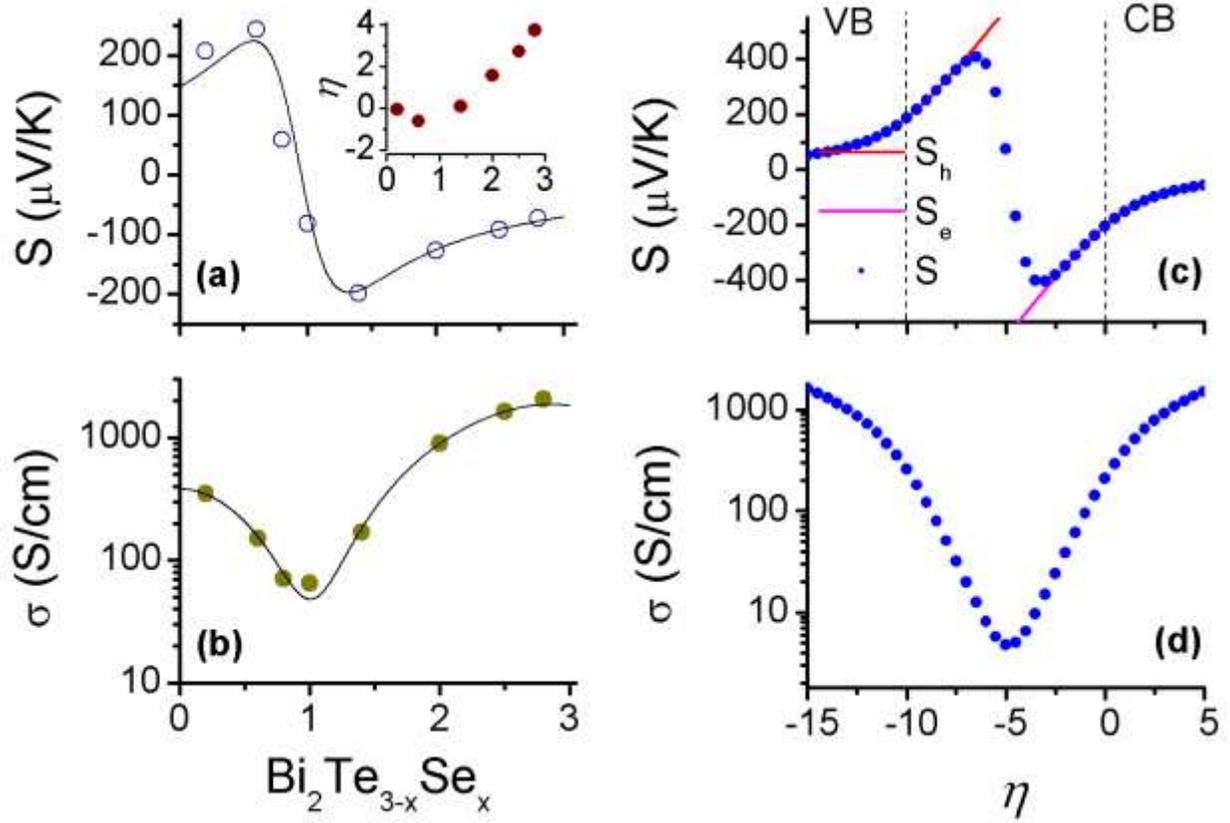

**FIG. 2** (Color online) Variation of the experimental Seebeck coefficient (a) and electrical conductivity (b) against Se content. The compositional dependence of the reduced chemical potential is shown in the inset of Fig. 2a. Solid lines are guides to eye. Calculated total Seebeck coefficient (c) and total electrical conductivity $\sigma = \sigma_h + \sigma_e$ (d) as a function of the reduced chemical potential for the $Bi_2Te_2Se$ material. At the p- to n- crossover the conductivity is minimized. For the calculations the electron effective mass and the electron mobility were taken $m_e^* = 1 m_o$ and $\mu_e = 100$ cm$^2$/Vs.

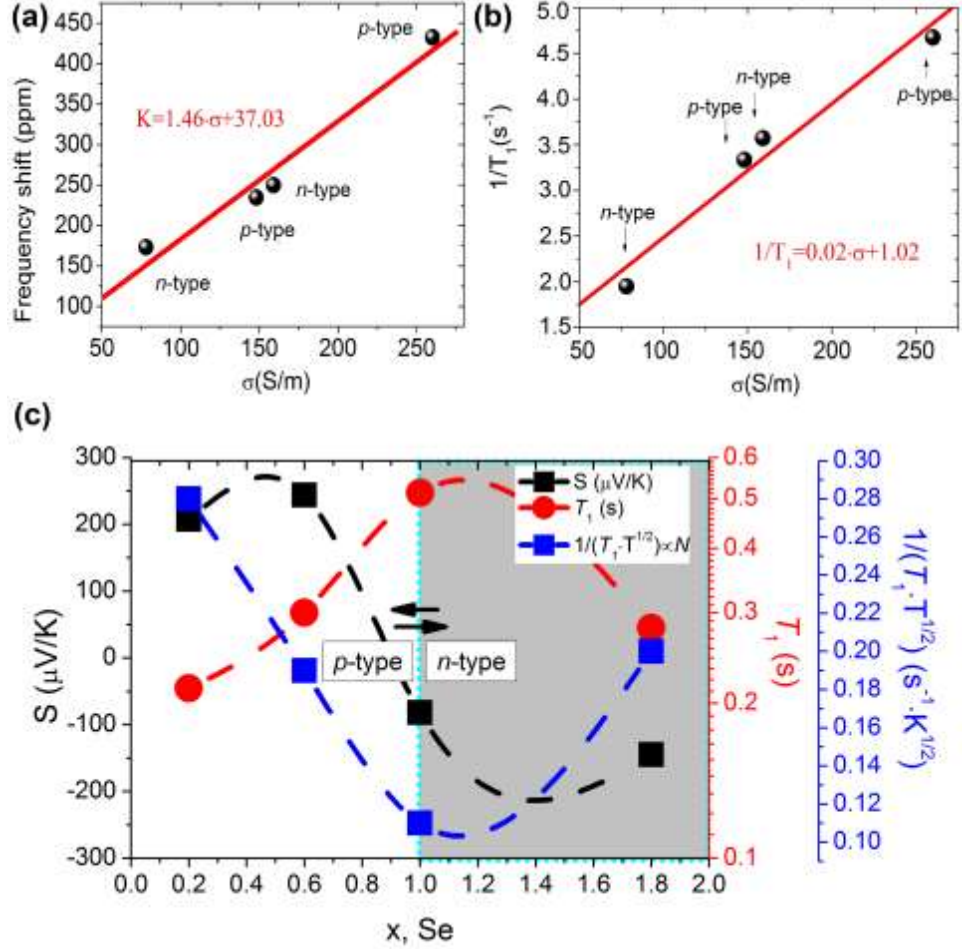

**FIG. 3** (Color online) (a) $^{125}$Te NMR shift versus conductivity. (b) $^{125}$Te spin lattice relaxation rate ($T_1^{-1}$) versus conductivity. The linear relationships indicate that the system follows a mixed conduction mechanism. (c) $^{125}$Te spin lattice relaxation time ($T_1$) and Seebeck coefficient (*S*) versus Se content. The NMR relaxation time accompanied by the Seebeck coefficient and the $1/(T_1 \cdot T^{1/2})$ product intersect at *x*=1. Dashed lines are guides to the eye.